\documentclass[article]{elsarticle}

\usepackage{mathtools}
\usepackage{amssymb}
\usepackage{lineno,hyperref}
\usepackage{graphicx}
\usepackage{subcaption}
\usepackage{float}
\modulolinenumbers[5]
\journal{Composites Part A: Applied Science and Manufacturing}
\modulolinenumbers[5]
\usepackage{textcomp}
\usepackage{amssymb}  
\usepackage{bm}       
\usepackage{booktabs} 
\usepackage{dcolumn}  
\usepackage{multirow} 
\usepackage{graphicx} 
\usepackage[printonlyused]{acronym}
\usepackage{mathtools}
\usepackage{lineno,hyperref}
\usepackage{subcaption}
\usepackage{float}
\usepackage{ragged2e}
\usepackage[usenames, dvipsnames]{color}
\graphicspath{{figures/}} 

\begin{document}
	
	\begin{frontmatter}
		
		\title{Revisiting the Deformation-Induced Damage in Filled Elastomers: Effect of Network Polydispersity}
		
		\author{Mohammad Tehrani}
		\author{Mohammad Hossein Moshaei}
		\author{Alireza Sarvestani\corref{cor1}}
		\ead{sarvesta@ohio.edu}
		\cortext[cor1]{Corresponding author}
		
		\address{Department of Mechanical Engineering, Ohio University, Athens OH 45701, USA}

\begin{abstract}
A priori assumption in micromechanical analysis of polymeric networks is that the constitutive polymer strands are of equal length. Monodisperse distribution of strands, however, is merely a simplifying assumption. In this paper, we relax this assumption and consider a vulcanized network with a broad distribution of strand length. In the light of this model, we predict the damage initiation and stress-stretch dependency in a filled polymer network with random internal structures. The degradation of network mechanical behavior is assumed to be controlled by the adhesive failure of the strands adsorbed to the filler surface. We show that the short adsorbed strands are the culprits for damage initiation and their finite extensibility is a key determinant of mechanical strength.
\end{abstract}

\begin{keyword}
	rubber composites, chain length distribution, damage initiation, polydispersity
\end{keyword}

\end{frontmatter}

\section{Introduction}
\noindent Small filler particles like carbon black, silica, and clay are often compounded with rubbers to improve their mechanical properties, including stiffness, abrasion resistance, tenacity, and durability \cite{hamed2000reinforcement,medalia1987effect,waddell1996use}. The fillers are also the major contributors to the damage nucleation and underlie the stress and strain softening mechanisms in filled elastomers. History dependence, or the \textquotedblleft Mullins effect\textquotedblright, is a particular feature of the mechanical response of filled elastomers in which the material shows hysteresis during quasi-static loading and softens with the history of loading. Another prominent example is the amplitude dependence of viscoelastic moduli of filled rubbers. Elastomers generally show linear viscoelastic properties in strain amplitudes up to 20 \cite{chazeau2000modulus}. The storage modulus of elastomers loaded with solid particles, however, shows a large drop with increasing strain. This strain-softening phenomenon is often referred to as the \textquotedblleft Payne effect\textquotedblright \cite{payne1962dynamic1,payne1962dynamic2,payne1965reinforcement}.

\bigskip
\noindent Great interest has been kindled in understanding the origin of damage initiation in mechanical behavior of filled elastomers due to its great practical importance in tire industry \cite{clark1978rolling,leblanc2009filled}. Over the past decades, different micro-mechanical and phenomenological mechanisms are proposed to explain the mechanism of damage nucleation and growth in filled elastomers \cite{leblanc2002rubber}. As initially noted by Payne and Kraus \cite{kraus1984mechanical,payne1972effect}, degradation of mechanical properties may arise from disruption of the agglomerated particles. Strong inter-particle interaction among surface active particles, like silica, leads to formation of disorderly grown aggregates with fractal structure, ranging from 10 to 100 nm in size \cite{kluppel2003role}. Increasing the filler content beyond the percolation threshold creates filler networks at larger scales within the matrix. Large strain perturbations deform and eventually disrupt the filler networks and introduce strong nonlinearity in the mechanical behavior of rubber composites \cite{heinrich2002recent}. 

\bigskip
\noindent In a different approach, the alteration of networks mechanical properties is ascribed to the nature of rubber-filler interactions. Polymer molecules generally show affinity for the surface of active particles. This is mediated either by chemical and strong physical bonds \cite{stockelhuber2011impact,leopoldes2004influence,ramier2007payne} or by disorder-induced localization of polymer onto the rough surface of particles \cite{vilgis1994disorder,vilgis2005time}. The structure of adsorbed polymer layer changes with applied deformation. Bueche \cite{bueche} argued that deformation of a filled rubber will break the highly stretched chains bridging the two adjacent fillers or tear them loose from the filler surface. Maier and G\"{o}ritz \cite{maier1996molecular} proposed a different mechanism, in which the affinity between fillers and polymer chains favors the establishment of stable and unstable bonds on the filler surface. The adhesion sites between the polymer chains and fillers are regarded as temporary and supplemental “crosslink points” contributing to the entropic elasticity. Unstable bonds have less drag-resistance and break under the elevated interfacial stresses \cite{raos2013pulling}. Large deformations promote the polymer disentanglement from filer surface causing the overall stiffness to drop. More recently, Jiang \cite{jiang2014effect} showed this Langmuir-type desorption of polymer chains could be due to the non-uniform stretching of chains in the matrix.

\bigskip
\noindent Most of the foregoing phenomenological descriptions have been integrated into micromechanical material models for filled rubbers (see Govindjee and Simo \cite{govindjee1991micro} and Heinrich and Kl\"{u}ppel \cite{heinrich2002recent} and references therein). The goal of present contribution is to add a new dimension to the constitutive modeling of bulk damage in filled vulcanizates by accounting for the random internal structure of rubber and the polydispersity of strands. Monodispersity of constitutive strands has been taken as a priori assumption in classical theory of rubber elasticity. The vulcanization of rubbers, however, is an inherently random process which likely results in formation of polydisperse networks. The importance of strand length distribution in mechanical behavior of vulcanizates was first recognized in the pioneering works of Bueche \cite{bueche}, Watson \cite{watson1953chain,watson1954chain}, and Gehman \cite{gehmanmolecular}. Models for mechanical behavior of polydisperse networks that take into account the statistical information of the strand length distribution are developed only recently \cite{itskovrubber,TehraniSar}. In this paper, we will examine the effect of polydispersity on elasticity and mechanical strength of filled vulcanizates with polydisperse structures. We develop a micromechanical model for initiation of bulk damage and show how the irregularity in network structure markedly affects the strength of filled rubbers.

\section{Model}
 
 \noindent Bueche \cite{bueche} and Watson \cite{watson1953chain,watson1954chain} first proposed a simple distribution function for strand length in a random polymer network. Accordingly, if ${n_{j}}$ represents the number of strands with ${j}$ statistical segments, the probability distribution of having a strand with ${j}$ statistical segments, $P(j)$, can be expressed as

\begin{equation} 
P(j)=\frac{1}{\overline{j}}e^{-j/\overline{j}}
\end{equation}

\noindent where $\overline{j}=\frac{1}{p}$ and ${p}$ shows the probability of a segment to be crosslinked. In deriving Eq. (1), it was assumed that the placement of crosslinks on the main chain is completely random, all segments have an equal chance ${p}$ to participate in vulcanization, and the segmental bindings during crosslinking are statistically independent events. $\overline{j}$ is a decay length (Figure 1) and represent the average strand length of the random network. 

\bigskip
\noindent Consider a random dispersion of rigid fillers, with volume fraction $\nu_{f}$ in an elastomeric matrix, with a strand length distribution that follows Eq. (1). The polymer strands close to the surface of particles may reversibly interact with the affine particle surface and establish labile bonds (Figure 2). This affinity leads to formation of a transition zone, with volume fraction $\nu_{f}$ around each particle in which segments of strands may adsorb (desorb) to (from) the particles. Upon adsorption, each strand with length ${j}$ is divided into two sub-chains with shorter length. For simplicity, we assume that the sub-chains have similar length and include ${j/2}$ statistical segments. 

\bigskip
\noindent Let $n_{j}^{(i)}(t)$ show the number density of the strands at time $t$ where the superscripts $j=a, f$, and $b$ refer to the adsorbed strands in the transition zone, free strands in the transition zone, and free strands in the bulk, respectively. The competitive adsorption/desorption of strands in the adhesion zone can be represented by the following kinetic equations

\begin{subequations}
	\label{eq:gen-dens-rel}
	\begin{align}
	&\frac{d n_{j}^{(a)}(t)}{d t}=-k_{r}n_{j}^{(a)}(t)+k_{f}n_{j}^{(f)}(t)\\
	&\frac{d n_{j}^{(f)}(t)}{d t}=-k_{f}n_{j}^{(f)}(t)+k_{r}n_{j}^{(a)}(t)\\
	&\frac{1}{2} n_{j}^{(a)}(t)+n_{j}^{(f)}(t)=n_{j}^{(b)}(t)
	\end{align}
\end{subequations}

\noindent where $k_{f}$ and $k_{r}$ stand for the forward and reverse rates of strand adhesion to the filler surface. Under a static or quasi-static loading condition, the concentrations of strands maintain their steady state values at

\begin{subequations}
	\label{eq:gen-quasi-rel}
	\begin{align}
	&n_{j}^{(a)}=\frac{\kappa n_{j}^{(b)}}{1+\frac{\kappa}{2}} \\
	&n_{j}^{(f)}=\frac{n_{j}^{(b)}}{1+\frac{\kappa}{2} } 
	\end{align}
\end{subequations}

\noindent where $\kappa=\frac{k_{f}}{k_{r}}$.

\bigskip
\noindent The composite system described above is subjected to an applied deformation gradient $\textbf{F}$. The average value of the local strain at each point in the matrix is somewhat larger than the applied strain, due to the hydrodynamic amplification \cite{bergstrom1999mechanical}. Following Mullins and Tobin \cite{mullins1965stress}, here we assume that the \textquotedblleft effective\textquotedblright stretch in the matrix reads 

\begin{equation}\label{eq:AmpStr}
\lambda_{e} =1+X(\lambda-1)
\end{equation}

\noindent where $\lambda$ is the applied stretch and 

\begin{equation}\label{eq:AmpStr}
X=1+2.5\nu_{f}+14.1\nu_{f}^{2}
\end{equation}

\noindent is the hydrodynamic correction factor. The end-to-end vector of each representative strand at the reference configuration is shown by $\textbf{R}^{(i)}_{0}$. After deformation, the strand finds a new configuration, shown by the end-to-end vector $\textbf{R}^{(i)}$. The strands are assumed to be flexible chains formed by $j$ freely joined statistical segments. Taking into account the Langevin effect, the stored energy in each strand, stretched by $\lambda_{e}$, is \cite{arruda1993three}

\begin{equation} 
w^{(i)}(\lambda_{e},j)=jk_{B}T\bigg (\frac{\lambda_{e}\beta}{\sqrt{j}}+\ln\frac{\beta}{\sinh\beta} \bigg)+w^{(j)}_{0}
\end{equation}

\noindent where $k_{B}T$ is the thermal energy, $w_{0}$ represents the deformation-independent part of the free energy, and 

\begin{equation} 
\beta=\pounds^{-1}\bigg (\frac{\lambda_{e}}{\sqrt{j}}\bigg ) \quad
\end{equation}

\noindent with $\pounds^{-1}$ being the inverse Langevin function.

\bigskip
\noindent The end point of the representative strand at the reference configuration is located on a sphere with radius of $R^{(i)}_{0}$, with spherical coordinates ${(R^{(i)}_{0},\theta_{0},\phi_{0})}$. If the medium is undeformed in the reference configuration, we may assume that the strands are randomly oriented in space before deformation. Thus, the number of strands with $j$ statistical segments whose endpoints falls in ${(R^{(i)}_{0},\theta_{0}+d\theta_{0},\phi_{0}+d\phi_{0})}$ is given by 

\begin{equation} 
dn^{(i)}_{j}=\frac{1}{4\pi}n^{(i)}_{j} sin\theta_{0} \ d\theta_{0} \ d\phi_{0}
\end{equation}

\noindent Hence, the total free energy of strands with $j$ statistical segments is

\begin{equation} 
W^{(i)}_{j}(\lambda_{e})= n^{(i)}_{j}\int\limits_{0}^{2\pi}\int\limits_{0}^{\pi} w^{(i)} \big (\lambda_{e},j \big ) \  \sin\theta_{0} \ d\theta_{0} \ d\phi_{0}
\end{equation}

\noindent The stretch along an arbitrary direction can be expressed in the reference configuration and in terms of the macroscopic principal stretches, $\lambda_{i}$, as

\begin{equation}
\lambda^{2} (\theta_{0},\phi_{0})=(\lambda_{1}  \sin\theta_{0}  \cos\phi_{0})^{2}+(\lambda_{2}  \sin\theta_{0}   \sin\phi_{0})^{2}+(\lambda_{3}  \cos\theta_{0})^{2}
\end{equation}

\noindent The total free energy of the entire population of chains, $W^{(i)}\big (\lambda_{e})$ is the summation of contribution of individual chains. That is, $W^{(i)}(\lambda_{e})=\sum_{j}^{}W^{(i)}_{j}(\lambda_{e})$. Considering a continuous distribution for polydisperse chains, as shown by Eq. (1), the summation can be replaced by an integral. For the bulk strands, it leads to

\begin{equation}
W^{(b)}(\lambda_{e})=\mu\int\limits_{0}^{2\pi}\int\limits_{0}^{\pi}\int\limits_{1}^{\infty} \ P^{(b)}(j) \ w \big (\lambda_{e},j \big )  \sin\theta_{0} \ dj \ d\theta_{0} \ d\phi_{0}
\end{equation}

\noindent where 

\begin{equation} 
P^{(b)}(j)=\frac{1}{\overline{j}^{(b)}}e^{-j/\overline{j}^{(b)}}
\end{equation}

\noindent and $\mu=\frac{\sum\sum_{j}^{} n_{j}}{4\pi}$. Similar equations can be derived for the free and adsorbed chains within the transition zone

\begin{subequations}
	\label{eq:gen-quasi-rel}
	\begin{align}
	&W^{(f)}(\lambda_{e})=\mu\int\limits_{0}^{2\pi}\int\limits_{0}^{\pi}\int\limits_{1}^{\infty} \ P^{(f)}(j) \ w \big (\lambda_{e},j \big )  \sin\theta_{0} \ dj \ d\theta_{0} \ d\phi_{0}\\
	&W^{(a)}(\lambda_{e})=\mu\int\limits_{0}^{2\pi}\int\limits_{0}^{\pi}\int\limits_{1}^{\infty} \ P^{(a)}(j) \ w \big (\lambda_{e},j \big )  \sin\theta_{0} \ dj \ d\theta_{0} \ d\phi_{0}
	\end{align}
\end{subequations}

\noindent $P^{(f)}(j)$ and $P^{(a)}(j)$ in Eqs. 13(a,b) are defined by the following ratios 

\begin{subequations}
	\label{eq:gen-quasi-rel}
	\begin{align}
	P^{(f)}(j)=\frac{n^{(f)}_{j}}{\sum_{j}^{}n_{j}}\\
	P^{(a)}(j)=\frac{n^{(a)}_{j}}{\sum_{j}^{}n_{j}}
\end{align}
\end{subequations}

\noindent Note that unlike $P^{(f)}(j)$ these two quantities do not represent the probability distribution of the free and adsorbed chains. They are simply the ratio of population free and adsorbed strands with $j$ segments within the transition zone to the number of total strands.

\bigskip
\noindent Using Eq. (4), the average entropic force developed in a strand with $j$ segments can be obtained as

\begin{equation} 
f_{j}(\lambda_{e})= \frac{k_{B}T\beta}{l}
\end{equation}

\noindent where $l$ is a characteristic length of a statistical segment. Eq. (15) accounts for the finite extensibility of strands and thus diverges as the stretch approaches the ultimate value of $\lambda_{lock}=\sqrt{j}$ \cite{arruda1993three}. Since the network is assumed to move affinely, shorter strands experience larger entropic tension even under a small macroscopic stretch. These large entropic forces lead to microscopic degradation of the filled network. The entropic tension of strands shortens the lifetime of the physical bonds. If tension is strong enough, free or bulk strands break at backbone or cleave at a crosslink and become elastically inactive. To take the deformation induced network alteration into account, we follow the method proposed by Itskov and Knyazeva \cite{itskovrubber} and Tehrani and Sarvestani \cite{TehraniSar} and replace the lower limit of the first integral in Eq. (11) and (13) with the shortest strand that can survive the macroscopic stretch. The dissociation energy of labile or covalent bonds is modeled using the Morse pair-potentials \cite{crist1984polymer,dal2009micro}

\begin{equation} \label{eq:LeghGe}
U^{(i)}(r)=U^{(i)}_{0}\Big(1-\textrm{exp}\big[-\alpha^{(i)}(r^{(i)}-r^{(i)}_{0})]\Big)^2
\end{equation}

\noindent where $U^{(i)}_{0}$ is the dissociation energy and $\alpha^{(i)}$ is a constant that determines bonds elasticity. $r^{(i)}$ and $r^{(i)}_{0}$ show the deformed and equilibrium length of a bond, respectively. The dissociation or rupture of strands occurs when the maximum entropic force developed in a strand reaches the critical value of $\frac{\alpha U^{(i)}_{0}}{2}$. Using Eq. (15), we can find $j^{(i)}_{min}$, the number of statistical segments of the shortest elastically active strand that survives the effective macroscopic stretch $\lambda_{e}$ as

\begin{equation} \label{eq:LeghGe}
j^{(i)}_{min}(\lambda_{e})=\frac{\lambda^{2}_{e}(\theta_{0},\phi_{0})}{\xi^{(i)}}
\end{equation}

where 

\begin{equation} 
\frac{1}{\xi^{(i)}}=\frac{3(3+\sqrt{4\gamma^{(i)}+9})}{2(\gamma^{(i)})^{2}}+1 \qquad , \quad \gamma^{(i)}=\frac{\alpha^{(i)} U^{(i)}_{0} l }{2k_{B}T}
\end{equation}

\noindent Note that $\xi^{(b)}=\xi^{(f)}$. Replacing the lower limit of the first integral in Eq. (11) and (13) with $j^{(i)}_{min}$, we obtain

\begin{equation} 
W^{(i)}(\lambda_{e})=\mu\int\limits_{0}^{2\pi}\int\limits_{0}^{\pi}\int\limits_{j^{(i)}_{min}(\lambda_{e})}^{\infty} \ P^{(i)}(j) \ w(\lambda_{e},j) \ \sin\theta_{0} \ dj \ d\theta_{0} \  d\phi_{0}
\end{equation}

\noindent The detachment of adsorbed strands changes the balance of their steady state density in the transition zone. We consider this effect using a force-dependent unbinding rate, resulted from thermal fluctuation theory \cite{zhurkov1966kinetic}

\begin{equation} 
k_{r}=k^{0}_{r}\textrm{exp}\big[f_{j}\delta\diagup k_{B}T\big]
\end{equation}

\noindent where $k^{0}_{r}$ is a constant and $\delta$ is an activation length.

\bigskip
\noindent The average of total strain energy stored in the matrix can be represented as 

\begin{equation} 
W=\upsilon_{b} W^{(b)}+\upsilon_{t}(W^{(f)}+W^{(a)})
\end{equation}

\noindent The stresses produced in the incompressible network can be derived from the strain energy density, using the spectral decomposition theorem

\begin{equation} 
\boldsymbol{\sigma}= \lambda_{(k)} \frac{\partial W}{\partial\lambda_{(k)}} (\boldsymbol{n}^{(k)}\otimes \boldsymbol{n}^{(k)})  
\end{equation} \label{eq:DefStress}

\noindent where $\lambda_{(k)}$ and $\textbf{n}^{(k)}$ are the eigenvalues and eigenvectors of applied deformation tensor, respectively.

\section{Results}
\noindent In this section, we present some numerical examples illustrating the effect of polydispersity on the overall static behavior of a filled network. We explicitly focus on the effect of average length index ($\overline{j}$), representing the randomness in network structure, and the strength parameter ($\xi^{(i)}$). To evaluate the integrals appearing in Eqs. (19), we used the so-called \textquotedblleft Puso's approximation\textquotedblright for the inverse Langevin function \cite{puso1994mechanistic}

\begin{equation}
\pounds^{-1}(\beta)\approx\frac{3\beta}{1-\beta^{3}}
\end{equation}

 \noindent and the resulting integrals were calculated using MATLAB. The results are expressed in terms of non-dimensional parameters $\overline{\kappa}^{0}=\frac{k^{0}_{r}}{k_{f}}$ and $\overline{\delta}=\frac{\delta}{l}$ satisfying
 
\begin{equation}
\overline{\kappa}=\frac{k_{r}}{k_{f}}=\overline{\kappa}^{0}\textrm{exp}\big[\beta \overline{\delta}\big]
\end{equation}
 
  \noindent In the presented examples, the binding energy between polymer segments is assumed to be much stronger than the adsorption between strands and the filler surface; i.e. $\xi^{(b)}\gg\xi^{(a)}$, unless otherwise is explicitly mentioned. Furthermore, the volume fraction of the transition zone is assumed to be equal to the volume fraction of fillers ($\nu_{t}=\nu{f})$.   
 
 \bigskip
 \noindent Figure 3 shows how the polydispersity affects the steady-state distribution of strands in the transition zone. We calculated $P^{(f)}(j)$ and $P^{(b)}(j)$ for a random network with $\overline{j}=80$ in the bulk. We chose a range of the $\overline{j}$ values comparable with the simulation results of Svaneborg et al. \cite{svaneborg2005disorder} for unfilled random networks. The sharp decrease in  distribution of bound strands indicates a large population of short adsorbed strands. Figure 4, shows the response of this system to a uniaxial deformation and compares it with filled networks at different $\overline{j}$ values. The effect of polydispersity on the ultimate strength is apparent. Comparatively, the networks with smaller $\overline{j}$ exhibit significantly lower strength. Due to the chosen value of $\xi^{(b)}$, the network degradation is essentially due to the rupture of labile bonds and adhesive failure of the adsorbed strands. To observe the effect of polydispersity more clearly, Figure 4 also includes the stress-stretch dependency in a monodisperse network reinforced with the same volume fraction of fillers, where all strands contain 80 statistical segments. The drastic difference between the behavior of monodisperse and polydisperse networks confirms that the network alteration is indeed induced by the structural polydispersity. The deformation-induces desorption is driven by the finite extensibility and non-Gaussian behavior of short strands in the transition zone. The entropic tension developed in shorter strands increases concurrent with macroscopic deformation and leads to desorption of a large population of short strands. In the presented model, the progressive desorption of strands is taken into account by evolution of $j^{(a)}_{min}$ with $\lambda$ (Figure 5).
 
\bigskip
\noindent Figure 6 shows the effect of relative values of strength parameters $\xi^{(a)}$ and $\xi^{(b)}$ on the variation of overall stress in the filled network. If the cohesive energy between strand monomers is strong  (i.e., $\xi^{(b)} \rightarrow 1$) increasing the energetic affinity between strands and fillers leads to enhancement of the ultimate strength of the composite. This is frequently reported in experimental measurements. For example, surface modification of carbon black with organic functional groups or fatty acids or increasing the hydrogen content on surface of carbon black are reported to moderately improve the tensile properties of filled SBR \cite{han2006effect,ghosh1997modified,ghosh1999reinforcing,ganguly2005effect}. In case of silica particles, the interaction between fillers and polymer can be enhanced by formation of covalent bonds. The surface chemistry of silica particles is different from carbon black, primarily due to superficial hydroxyl and geminal silanol groups \cite{Donnet1994}. Bifunctional organosilanes have been effectively used, particularly in tire industry, to facilitate the covalent bonding between silica particles and rubber molecules \cite{byers1998silane,hashim1998effect}. It has been proposed that if rubber-particle adhesion is stronger than the cohesion of rubber matrix, the damage may occur in the bulk instead of the interface \cite{suzuki2005effects}. Our model, qualitatively predicts such possibility when $\xi^{(a)} > \xi^{(b)}$, in which case the softening in mechanical response is essentially due to the alteration of bulk strands (for example by chain scission). 

\bigskip
\noindent Hysteresis is another characteristic aspect of the mechanical response of filled rubbers. Figure 7 illusrates this chiefly with reference to the effect of network polydispersity. Filled networks with lower $\overline{j}$ exhibit a more pronounced hsyteresis during a qusi-static deformation in accordance with the physics outlined here. This can be ascribed to the energy dissipation due to tearing loose of a large population of short strands initially adsorbed to the surface of filler particles. As expected, the volume fraction of particles markedly affects the mechanical behavior of the network. The areas of hystersis loops diretly correlate with the filler content (Figure 8 (a)). Furthermore, A larger filler content improves the strength but has a compromising effect on the deformation at break (Figure 8 (b)). Enhanced strength can be attributed to the amplification of stretch field by hydrodynamic correction factor $X$ and the stress production by highly stretched short strands in the transition zone. Increasing deformation is followed by desorption of short strands, which controls the onset of softening and reduces the deformation at break with increasing the filler content.

\bigskip
\noindent Finally, to examine the predictive ability of the proposed model, we compared the model predictions with a collection of experimental data represented by Meissner and Mat\v{e}jka \cite{meissner2001description} (Figure 9). They collected a series of experimental data presenting the stress-stretch dependency of unfilled and filler rubbers subjected to uniaxial deformation. We chose the result for SBR reinforced with different fractions of carbon blacks. SBR is an amorphous copolymer which does not undergo crystallization on stretching. The curves are shown in the so-called Mooney-Rivlin coordinates. The model parameters are obtained by fitting to the experimental data of neat SBR. The calibrated model was used to predict the response of filled networks with 20\% and 30\% carbon black, without changing any other model parameter.

\section{Concluding Remarks}

\noindent In this contribution, we proposed a new mechanism that is sufficient to cause marked alteration in the structure of filled polymer networks and can potentially influence the damage initiation and control the softening of filled elastomers in response to deformation. The proposed theory is based on the assumption that internal structure of vulcanizates is random and the distribution of strand length is polydisperse in nature. After compounding filler particles into a rubber matrix, each particle interacts with several polymer strands each of different length. This energetic interaction forms a myriad of reversible attachments between strands and particle surface. It was proposed that the alteration in mechanical behavior results from desorption of these physical bonds. During deformation shorter adsorbed strands reach their maximum elongation first and then snap. The cohesive failure of short and highly elongated strands continues concurrently with applied deformation. The softening can thus be attributed to a significant reduction in density of adsorbed strands and progressive loss of friction between the filler and rubber matrix. It was shown that polydispersity contributes in the energy loss in a cycle of loading and unloading and effectively enhance the hysteresis. Although implemented for quisi-static loadings, the theory laid out here may well explain some of the main characteristic features of the Payne effect in filled rubbers subjected to dynamic deformations. Namely, it describes the direct correlation between storage modulus with the filler content at small deformations, the drop in storage modulus with increasing strain, and shifting the onset of nonlinearity to lower deformations with increasing the filler content. 

\bigskip
\noindent The model is built upon multiple simplifying assumptions. Probably the most tenuous assumption here is the affine motion of all strands. Furthermore, the Morse potential is only a simple representation of bonding along the backbone and adhesion to the filler surface. In spite of its simplicity, the model provides compelling numerical evidences to support the effects of polydispersity on mechanical performance of filled vulcanizates. The main challenge, however, is to directly measure the strand length distribution in randomly crosslinked networks. Long ago, Gehman \cite{gehmanmolecular} referred to the distribution of network strands as a \textquotedblleft mental concept\textquotedblright that cannot be directly measured. Precise observations are eluded primarily due to the insolubility of crosslinked networks. Indirect measurements such as relaxometry of stressed networks or measurement of post-swelling pressure only provided limited information about microstructural irregularities in polymer networks. Light scattering and small-angle X-ray and neutron scattering techniques, however, confirmed spatial fluctuation in crosslink density in networks on scales of 1-100 nm \cite{seiffert2016scattering}. Further developments along this line may eventually help us obtain a correct estimate of network chain distribution, a challenge that still exists today.

\section*{References}
\bibliography{mybibfile}
\newpage
\begin{figure}[H]
	\centering
	\includegraphics[width=\linewidth]{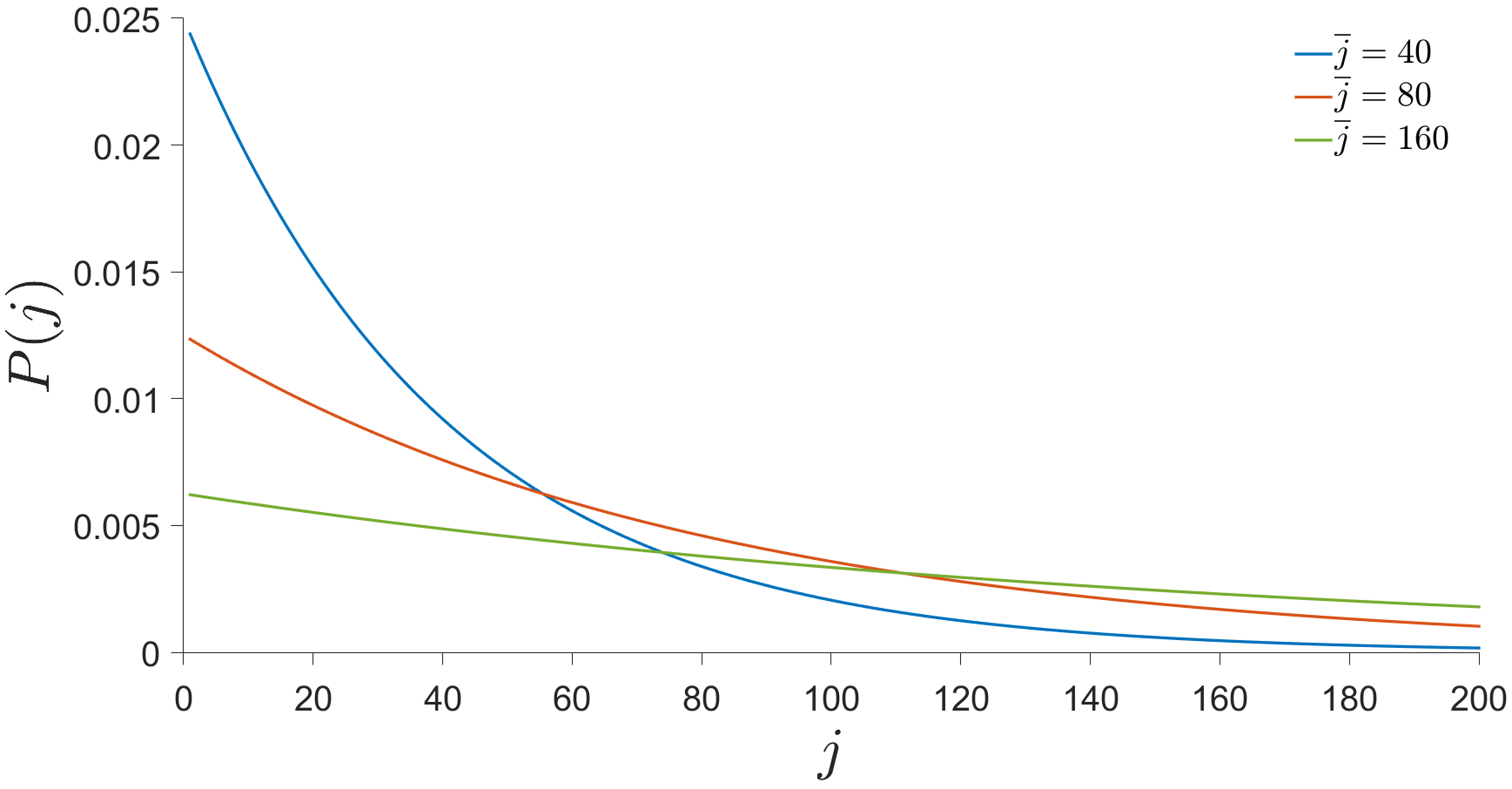}
	\caption{Probability distribution for strand length with different average values. Networks with lower $\overline{j}$ include a larger population of short strands.}
	\label{fig:configurations}
\end{figure}
\newpage
\begin{figure}[H]
	\centering
	\includegraphics[width=\linewidth]{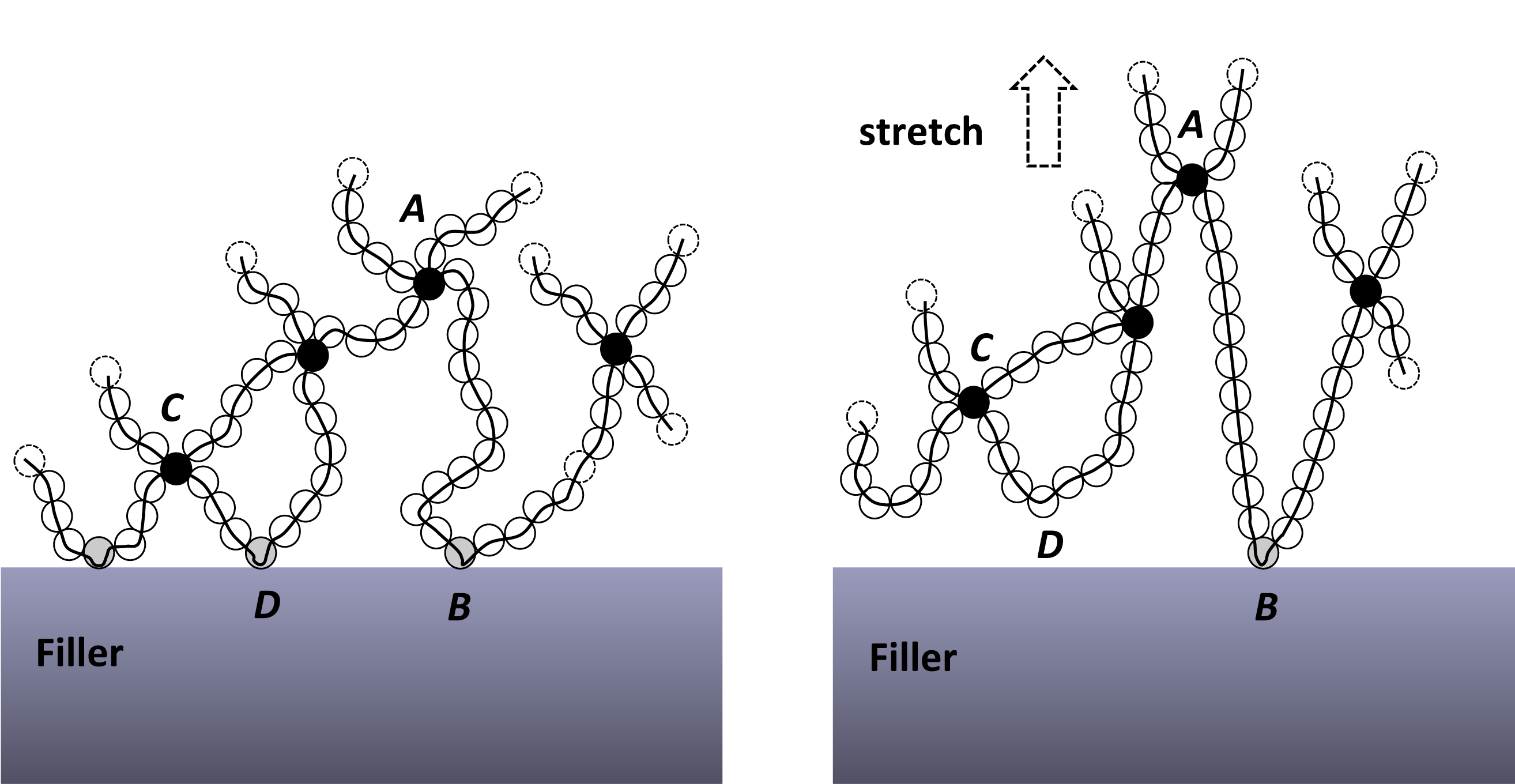}
	\caption{A polydisperse network close to the surface of a filler. The circles schematically show the statistical segments forming the polymer chains. (Left) Before application of macroscopic deformation, polymer strands may attach to the filler surface and form physical bonds at multiple points (B and D). AB and CD show two adsorbed strands with different lengths. (Right) As the system deforms, shorter strands (e.g., CD) desorb from the filler surface due to relatively larger entropic forces developed in them.}
	\label{fig:configurations}
\end{figure}
\newpage
\begin{figure}[H]
	\centering
	\includegraphics[width=\linewidth]{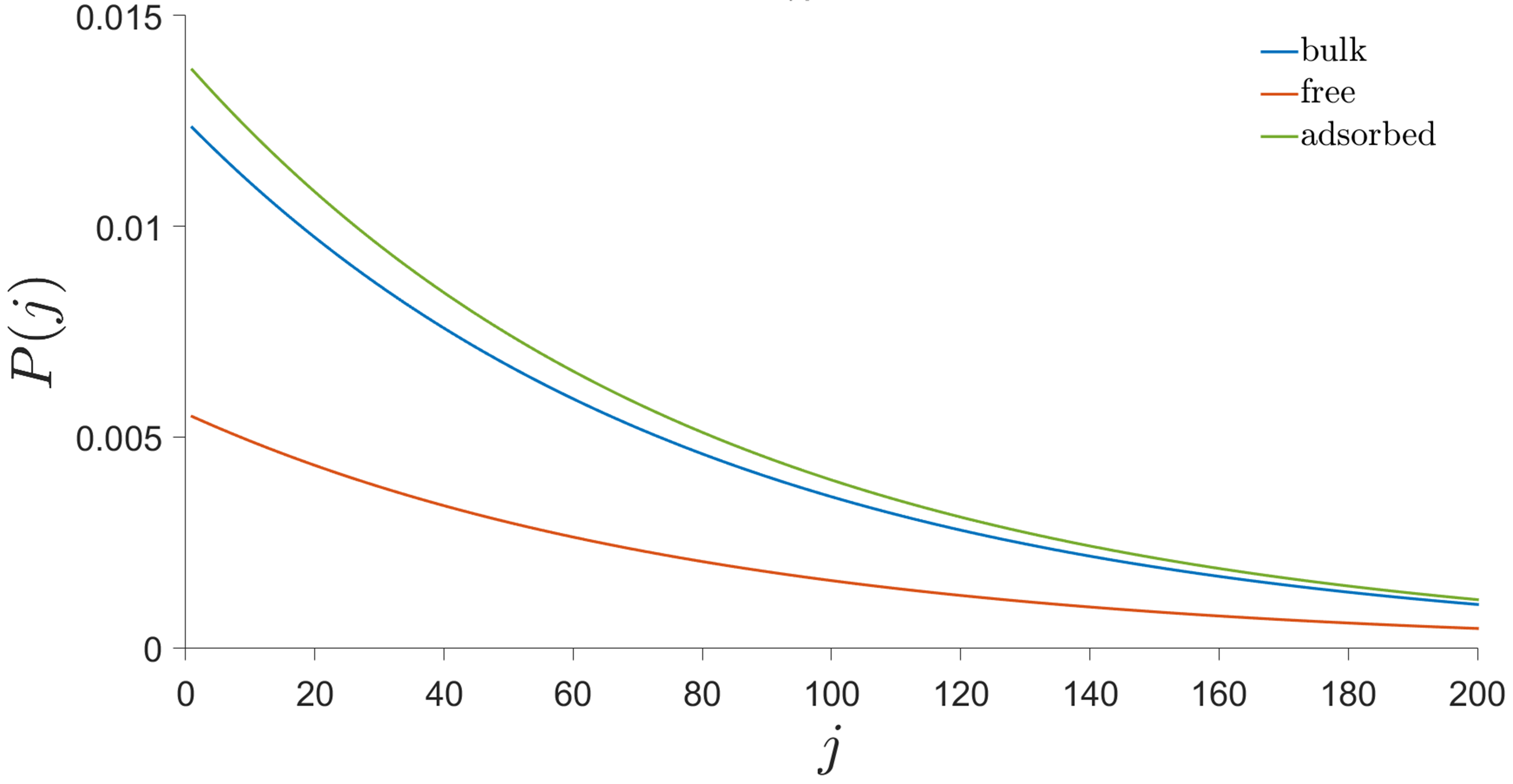}
	\caption{Variation of $P^{(i)}(j)$ functions for the adsorbed and free strands in the transition zone in comparison with that in the bulk ($\overline{j}=80$, $\nu_{f}=0.1$, $\overline{\kappa}^{0}=2.5$, $\overline{\delta}=0.2$).}
	\label{fig:configurations}
\end{figure}
\newpage
\begin{figure}[H]
	\centering
	\includegraphics[width=\linewidth]{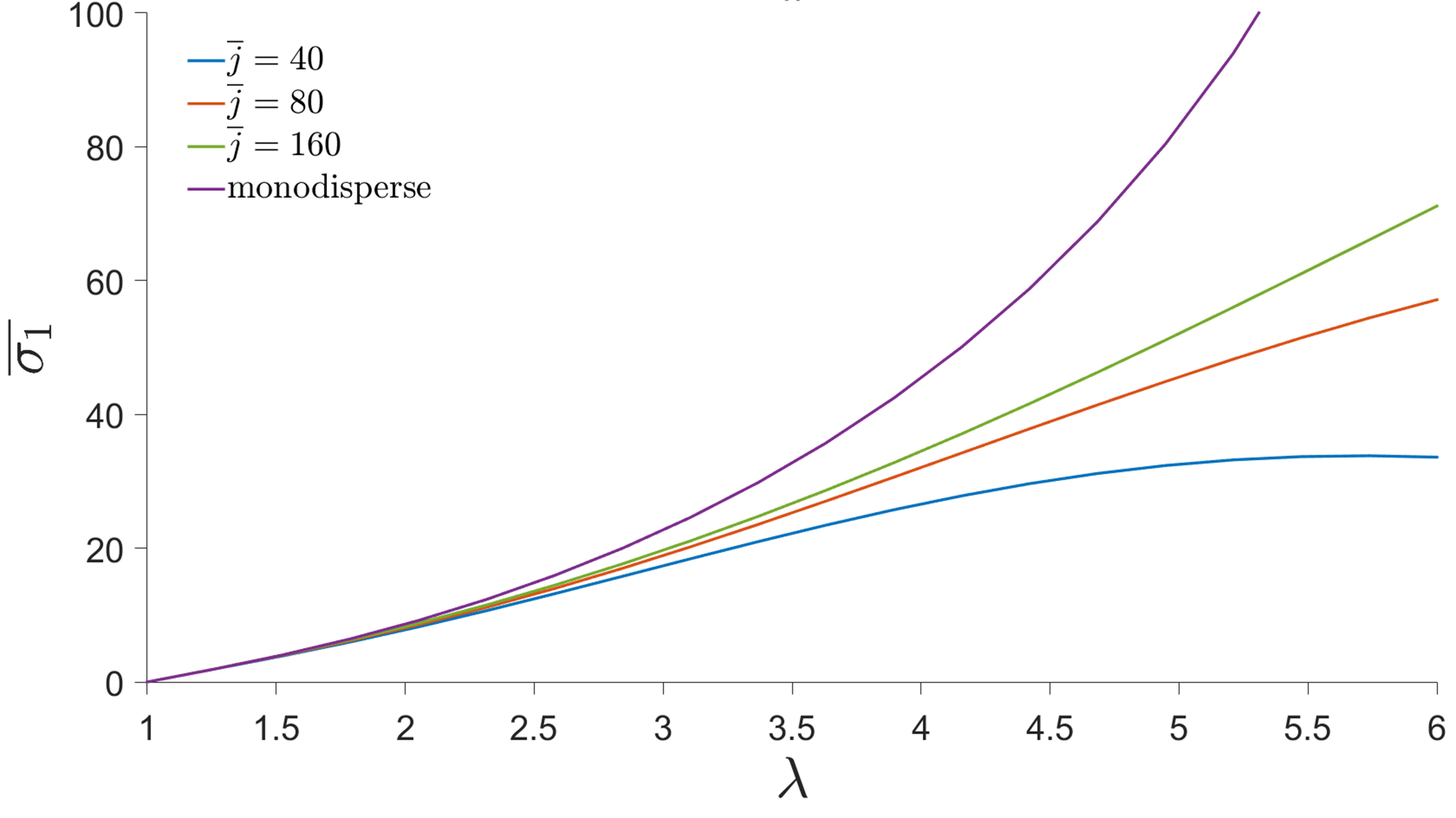}
	\caption{The effect of average strand length, $\overline{j}$, on the stress-stretch dependency in random networks subjected to uniaxial deformation. The monodisperse network is formed by strands with 80 statistical segments ($\overline{\sigma_{1}}=\sigma_{1}/\mu$, $\nu_{f}=0.1$, $\xi^{(b)}=\xi^{(f)}=0.99$, $\xi^{(a)}=0.2$, $\overline{\kappa}^{0}=2.5$, $\overline{\delta}=0.2$).}
	\label{fig:configurations}
\end{figure}
\newpage
\begin{figure}[H]
	\centering
	\includegraphics[width=\linewidth]{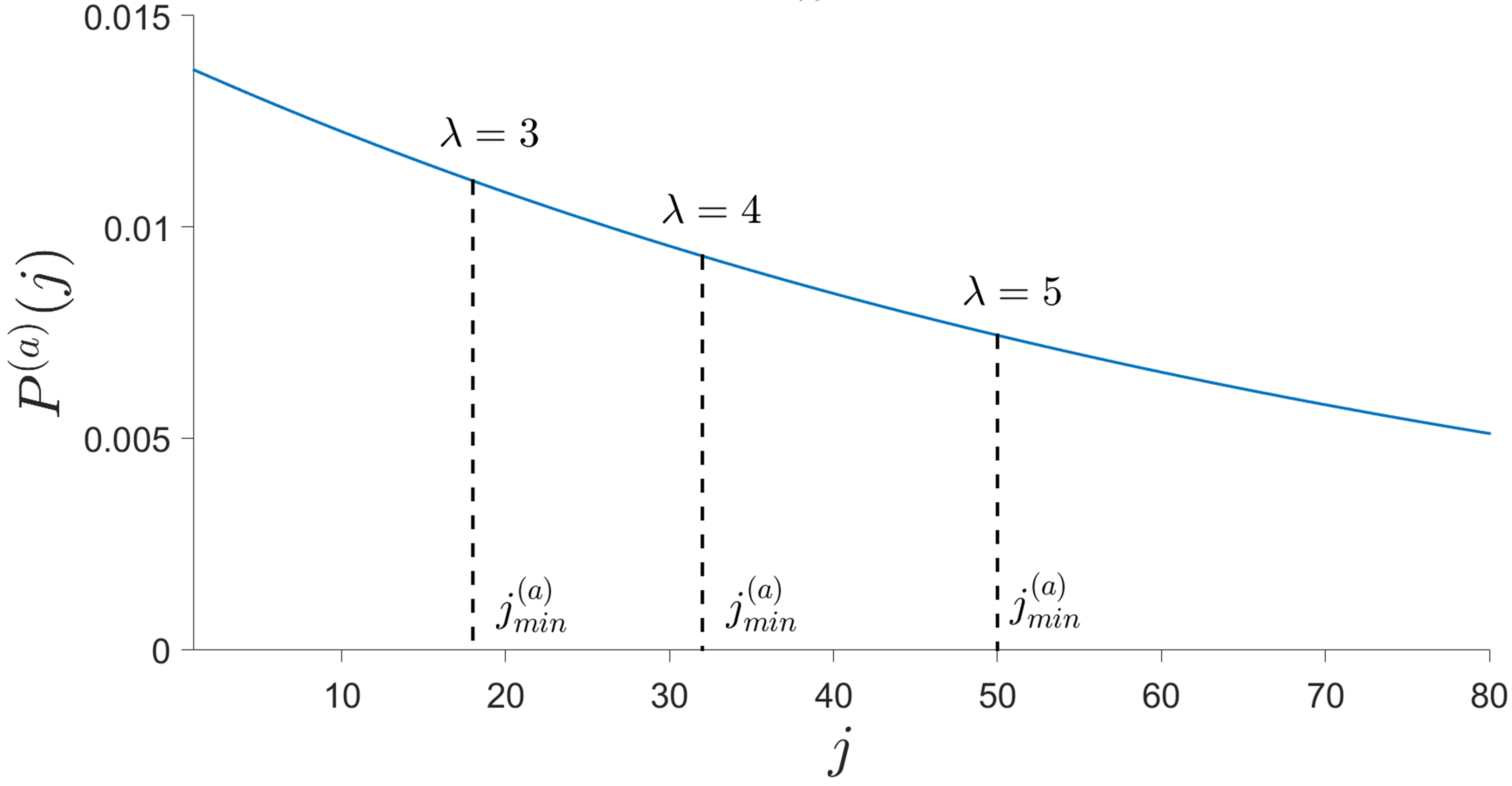}
	\caption{Evolution of $j^{(a)}_{min}$, the length of the shortest surviving adsorbed strand, with macroscopic stretch during a uniaxial deformation ($\overline{j}=80$, $\nu_{f}=0.1$, $\xi^{(b)}=\xi^{(f)}=0.99$, $\xi^{(a)}=0.2$, $\overline{\kappa}^{0}=2.5$, $\overline{\delta}=0.2$).}
	\label{fig:configurations}
\end{figure}
\newpage
\begin{figure}[H]
	\centering
	\includegraphics[width=\linewidth]{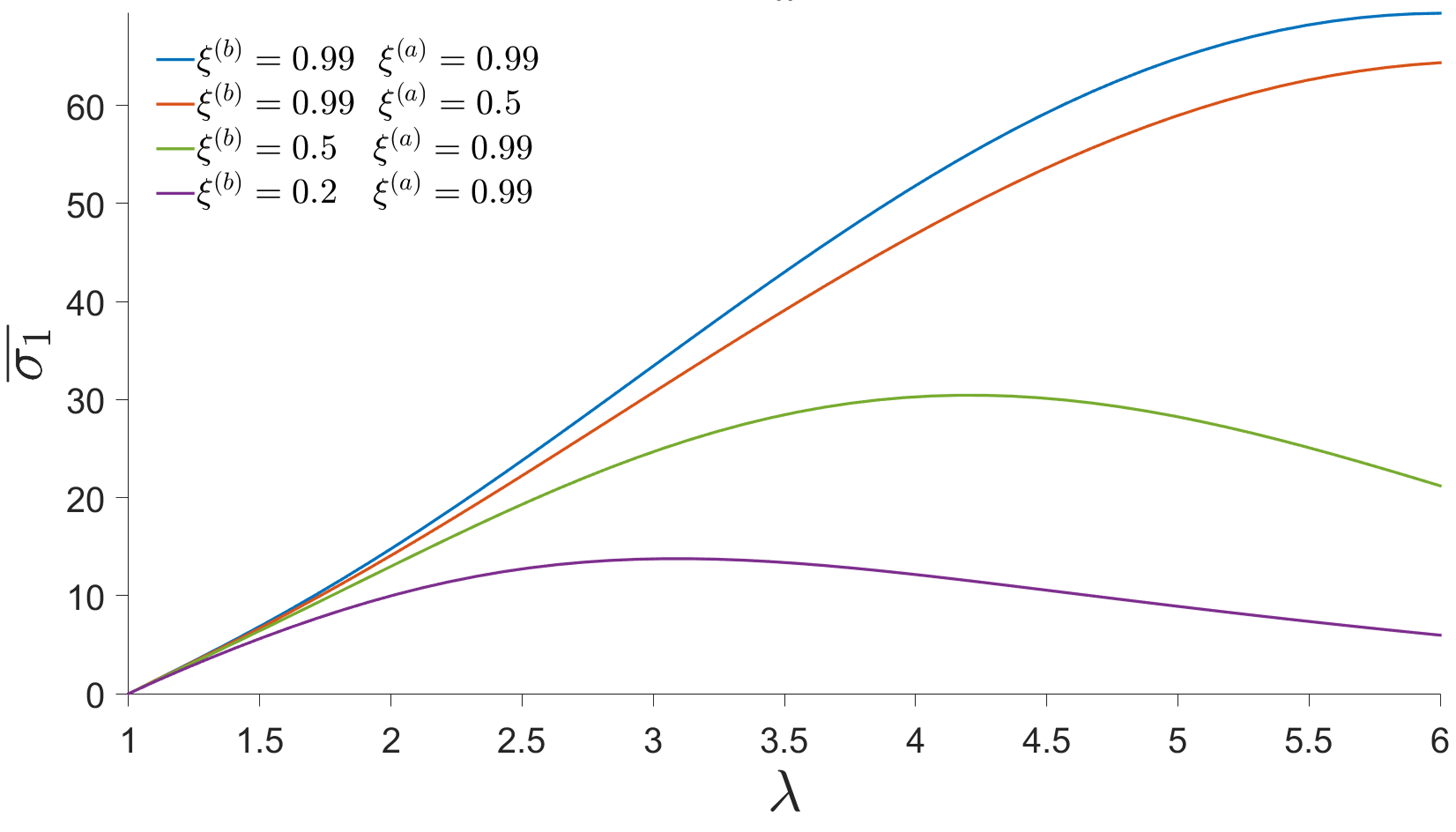}
	\caption{Effect of the strength parameter $\xi^{(i)}$ on the network strength subjected to a uniaxial deformation ($\overline{\sigma_{1}}=\sigma_{1}/\mu$, $\overline{j}=80$, $\nu_{f}=0.1$, $\overline{\kappa}^{0}=2.5$, $\overline{\delta}=0.2$).}
	\label{fig:configurations}
\end{figure}
\newpage
\begin{figure}[H]
	\centering
	\includegraphics[width=\linewidth]{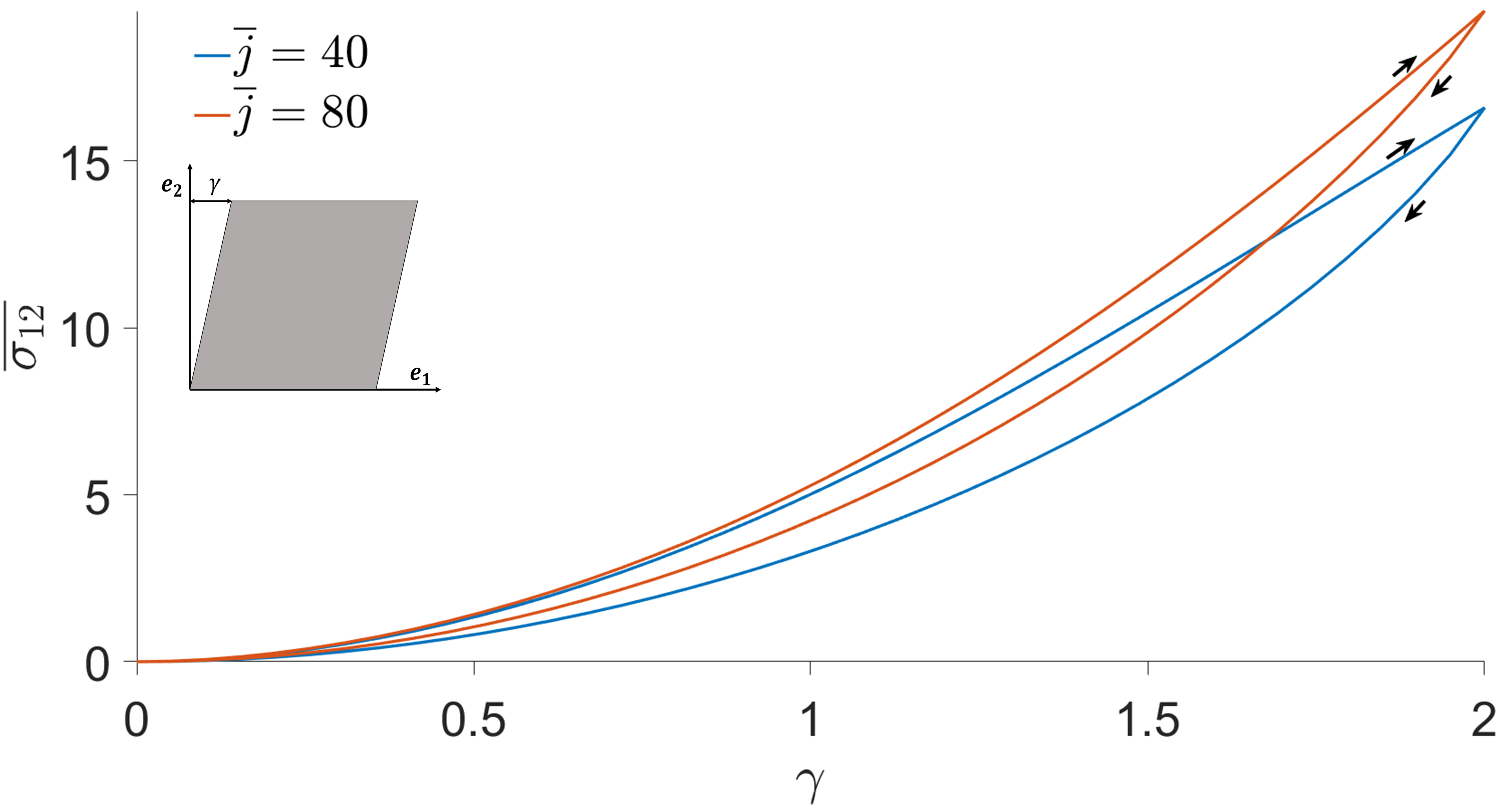}
	\caption{Effect of  filler volume fraction on the network strength subjected to a simple shear deformation ($\overline{\sigma_{12}}=\sigma_{12}/\mu$, $\overline{j}=80$, $\xi^{(b)}=\xi^{(f)}=0.99$, $\xi^{(a)}=0.2$, $\overline{\kappa}^{0}=2.5$, $\overline{\delta}=0.2$). The dashed lines show the predicted response of monodisperse networks whose strands are formed by 80 statistical segments.}
	\label{fig:configurations}
\end{figure}
\newpage
\begin{figure}[H]
	\centering
	\includegraphics[width=\linewidth]{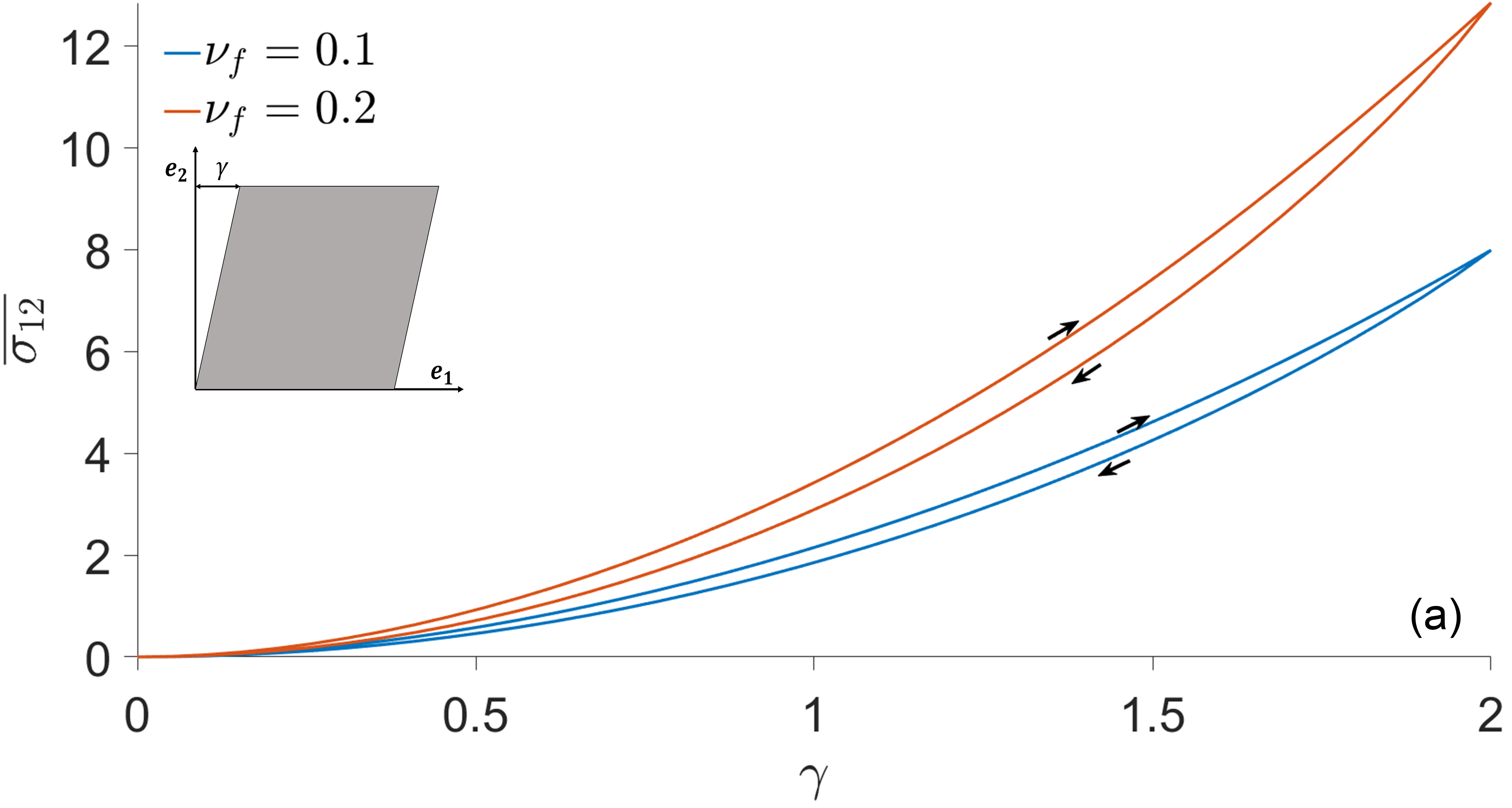}
	\caption{Effect of  filler volume fraction on the network strength subjected to a simple shear deformation ($\overline{\sigma_{12}}=\sigma_{12}/\mu$, $\overline{j}=80$, $\xi^{(b)}=\xi^{(f)}=0.99$, $\xi^{(a)}=0.2$, $\overline{\kappa}^{0}=2.5$, $\overline{\delta}=0.2$). The dashed lines show the predicted response of monodisperse networks whose strands are formed by 80 statistical segments.}
	\label{fig:configurations}
\end{figure}
\newpage
\begin{figure}[H]
	\centering
	\includegraphics[width=\linewidth]{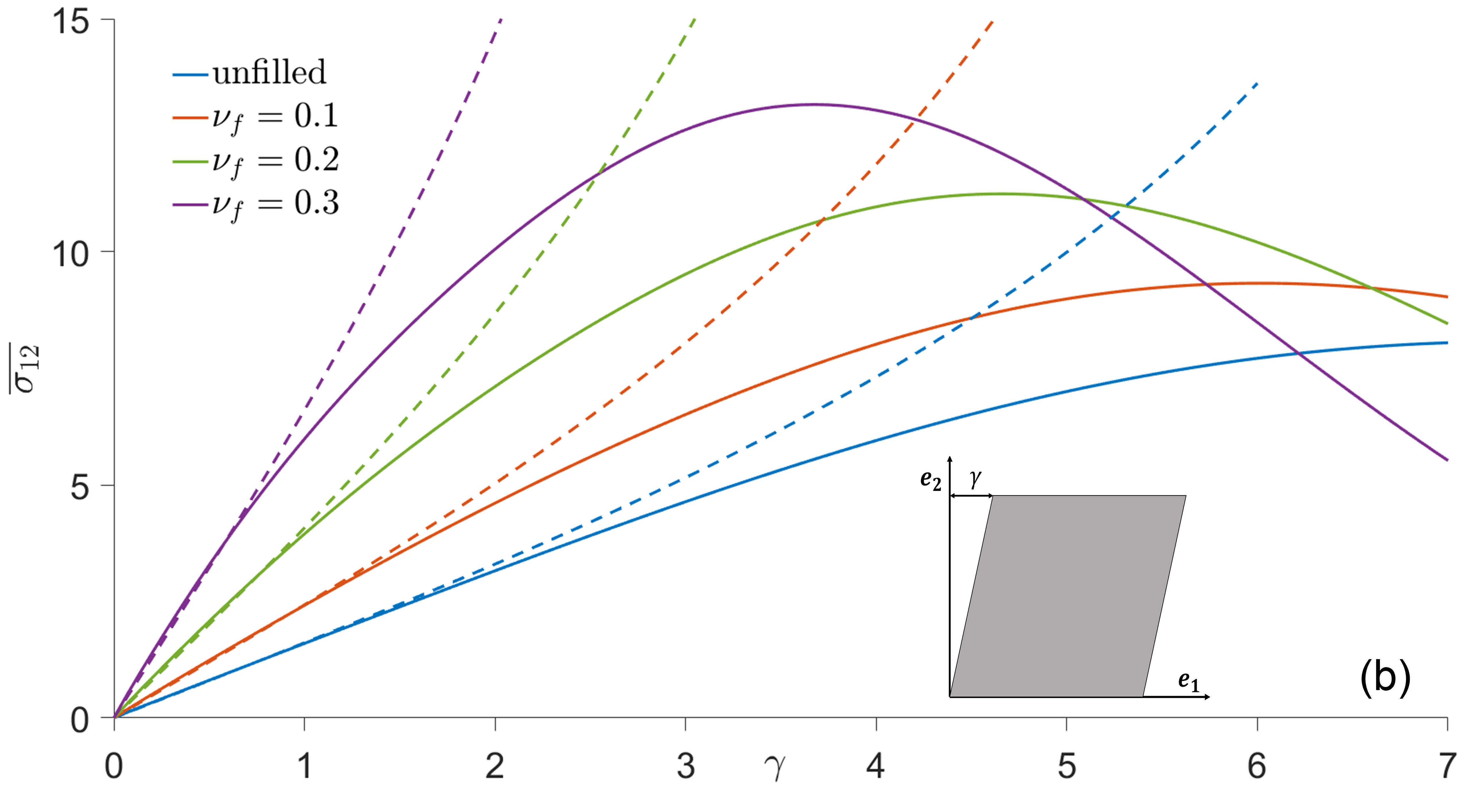}
	\caption{Effect of  filler volume fraction on the network strength subjected to a simple shear deformation ($\overline{\sigma_{12}}=\sigma_{12}/\mu$, $\overline{j}=80$, $\xi^{(b)}=\xi^{(f)}=0.99$, $\xi^{(a)}=0.2$, $\overline{\kappa}^{0}=2.5$, $\overline{\delta}=0.2$). The dashed lines show the predicted response of monodisperse networks whose strands are formed by 80 statistical segments.}
	\label{fig:configurations}
\end{figure}
\newpage
\begin{figure}[H]
	\centering
	\includegraphics[width=\linewidth]{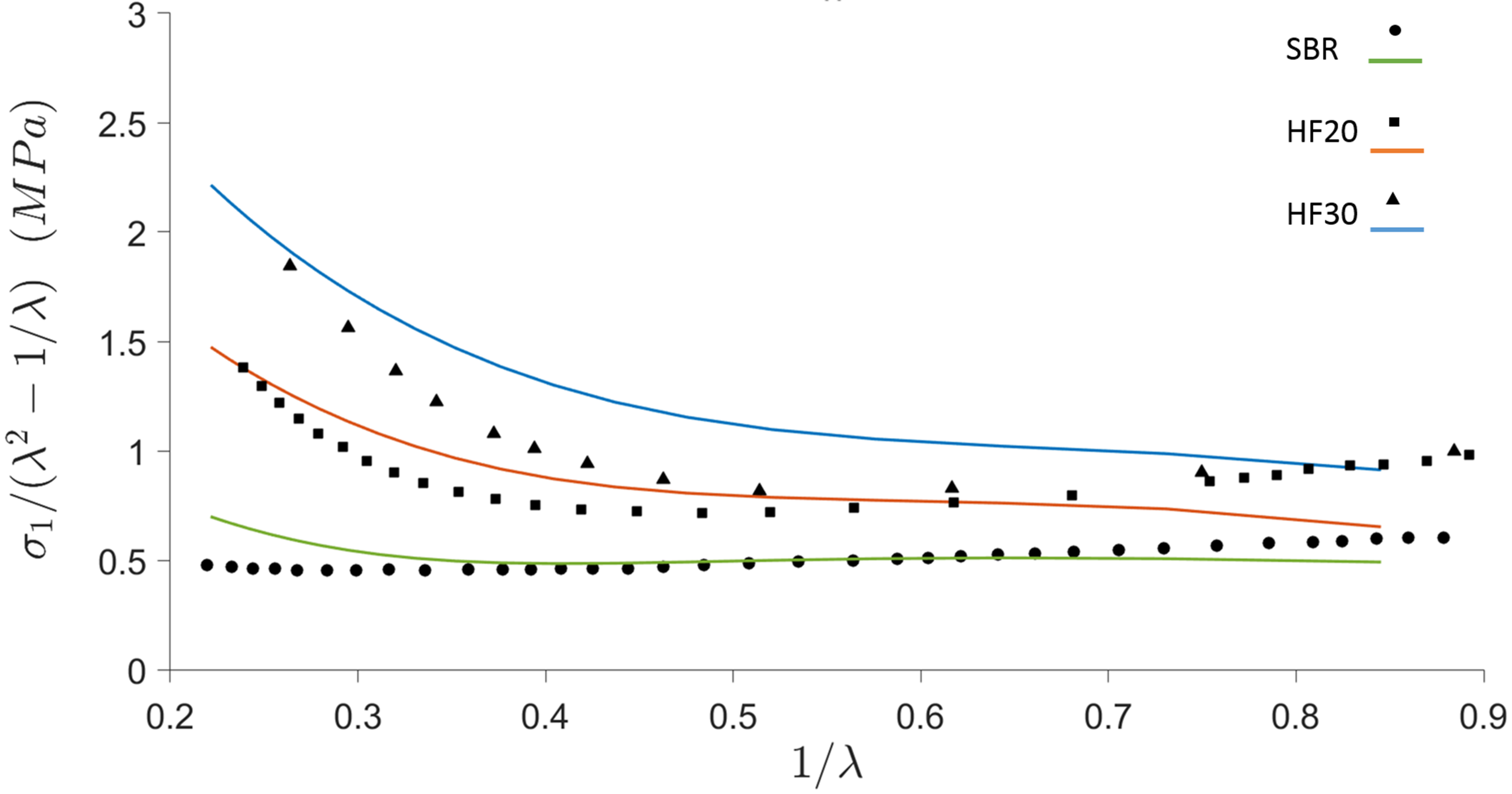}
	\caption{The comparison between proposed model (solid lines) and the experimental data (symbols) presented by Meissner and Mat\v{e}jka \cite{meissner2001description} for unfilled and filled SBR ($\mu=0.2875 \ m^{-3}$, $\overline{j}=160$, $\xi^{(b)}=\xi^{(f)}=0.99$, $\xi^{(a)}=0.1$, $\overline{\kappa}^{0}=2.5$, $\overline{\delta}=0.2$). The model was calibrated by fitting to the experimental data of neat SBR and was used to predict the response of filled networks (HF20, HF30) with no further calibration.}
	\label{fig:configurations}
\end{figure}
\end{document}